\documentclass[twocolumn]{aastex7}
\graphicspath{{figures/}}

\usepackage{xspace}

\newcommand{\Kepler}{\textit{Kepler}\xspace}
\newcommand{\TESS}{TESS\xspace}
\newcommand{\Gaia}{\textit{Gaia}\xspace}
\newcommand{\Tierras}{\textit{Tierras}\xspace}

\newcommand{\Exofast}{\texttt{EXOFASTv2}\space}
\newcommand{\rmfit}{\texttt{rmfit}\space}

\newcommand{\bjdtdb}{\ensuremath{\mathrm{BJD}_\mathrm{TDB}}\xspace}
\newcommand{\Rstar}{\ensuremath{R_{\star}}\xspace} 
\newcommand{\Mstar}{\ensuremath{M_{\star}}\xspace}
\newcommand{\Rjup}{\ensuremath{R_\mathrm{J}}\xspace} 
\newcommand{\Mjup}{\ensuremath{M_\mathrm{J}}\xspace}
\newcommand{\Rearth}{\ensuremath{R_\oplus}\xspace} 
\newcommand{\Mearth}{\ensuremath{M_\oplus}\xspace}

\newcommand{\Rp}{\ensuremath{R_p}\xspace}
\newcommand{\Mp}{\ensuremath{M_p}\xspace}

\newcommand{\Teff}{\ensuremath{T_\mathrm{eff}}\xspace}

\newcommand{\feh}{\ensuremath{\mathrm{[Fe/H]}}\xspace}
\newcommand{\vsini}{\ensuremath{v\sin{i_\star}}\xspace}
\newcommand{\veq}{\ensuremath{v_\mathrm{eq}}\xspace}
\newcommand{\veqsini}{\ensuremath{v_\mathrm{eq}\sin{i_\star}}\xspace}

\newcommand{\logrhk}{\ensuremath{\log{R^\prime_\mathrm{HK}}}\xspace}
\newcommand{\Prot}{\ensuremath{P_\mathrm{rot}}\xspace}
\newcommand{\istar}{\ensuremath{i_{\star}}\xspace}

\newcommand{\ms}{\ensuremath{\mathrm{m}\,\mathrm{s}^{-1}}\xspace}

\newcommand{\kms}{\ensuremath{\mathrm{km}\,\mathrm{s}^{-1}}\xspace}

\newcommand{\tierrasenddate}{19~July~2025}
\newcommand{\FitProt}{\ensuremath{26.4^{+0.9}_{-0.8}}}

\newcommand{\FitAge}{\ensuremath{{4.5}^{+2.1}_{-1.4}}}
\newcommand{\FitLambda}{\ensuremath{{81^\circ}^{+23^\circ}_{-22^\circ}}}
\newcommand{\FitPsi}{\ensuremath{{84.6^\circ}^{+7.1^\circ}_{-7.3^\circ}}}
\newcommand{\FitVsini}{0.46 \pm 0.11}
\newcommand{\FitIstar}{\ensuremath{{19.5^\circ}^{+5.1^\circ}_{-4.7^\circ}}}
\newcommand{\FitLambdaRMReloaded}{\ensuremath{{65^\circ}^{+32^\circ}_{-24^\circ}}}
\newcommand{\FitPsiRMReloaded}{\ensuremath{{85.9^\circ}^{+8.6^\circ}_{-9.2^\circ}}}
\newcommand{\FitVsiniRMReloaded}{\ensuremath{{0.42}^{+0.13}_{-0.11}}}

\begin{document}

\title{The Polar Orbit of TOI-2374\,b, a Planet in the Neptunian Ridge}

\author[0000-0001-7961-3907]{Samuel W. Yee}\altaffiliation{51 Pegasi b Fellow}
\affiliation{Center for Astrophysics $\vert$ Harvard \& Smithsonian, 60 Garden Street, Cambridge, MA 02138, USA}
\email[show]{samuel.yee@cfa.harvard.edu}

\author[0000-0003-2171-5083]{Patrick Tamburo}
\affiliation{Center for Astrophysics $\vert$ Harvard \& Smithsonian, 60 Garden Street, Cambridge, MA 02138, USA}
\email{patrick.tamburo@cfa.harvard.edu}

\author[0000-0001-7409-5688]{Gudmundur Stef{\'a}nsson}
\affiliation{Anton Pannekoek Institute for Astronomy, University of Amsterdam, 904 Science Park, Amsterdam, 1098 XH}
\email{g.k.stefansson@uva.nl}

\author[0000-0003-1361-985X]{Juliana Garc\'ia-Mej\'ia}\altaffiliation{51 Pegasi b Fellow}
\affiliation{Center for Astrophysics $\vert$ Harvard \& Smithsonian, 60 Garden Street, Cambridge, MA 02138, USA}
\affiliation{Kavli Institute for Astrophysics and Space Research, Massachusetts Institute of Technology, Cambridge, MA 02139, USA}
\email{jgarciam@mit.edu}

\author[0000-0002-9003-484X]{David Charbonneau}
\affiliation{Center for Astrophysics $\vert$ Harvard \& Smithsonian, 60 Garden Street, Cambridge, MA 02138, USA}
\email{dcharbonneau@cfa.harvard.edu}

\author[0000-0003-1464-9276]{Khalid~Barkaoui}
\affiliation{Instituto de Astrof\'isica de Canarias (IAC), E-38205 La Laguna, Tenerife, Spain}
\affiliation{Astrobiology Research Unit, University of Li\`ege, All\'ee du 6 ao\^ut, 19, 4000 Li\`ege (Sart-Tilman), Belgium}
\affiliation{Department of Earth, Atmospheric and Planetary Sciences, Massachusetts Institute of Technology, Cambridge, MA 02139, USA}
\email{Khalid.Barkaoui@uliege.be}

\author[0000-0001-6588-9574]{Karen A.\ Collins}
\affiliation{Center for Astrophysics $\vert$ Harvard \& Smithsonian, 60 Garden Street, Cambridge, MA 02138, USA}
\email{karen.collins@cfa.harvard.edu}

\author[0000-0001-8227-1020]{Richard~P.~Schwarz}
\affiliation{Center for Astrophysics $\vert$ Harvard \& Smithsonian, 60 Garden Street, Cambridge, MA 02138, USA}
\email{rpschwarz@comcast.net}

\author[0000-0001-8511-2981]{Norio Narita}
\affiliation{Komaba Institute for Science, The University of Tokyo, 3-8-1 Komaba, Meguro, Tokyo 153-8902, Japan}
\affiliation{Astrobiology Center, 2-21-1 Osawa, Mitaka, Tokyo 181-8588, Japan}
\affiliation{Instituto de Astrofisica de Canarias (IAC), 38205 La Laguna, Tenerife, Spain}
\email{Norio Narita <narita@g.ecc.u-tokyo.ac.jp>}

\author[0000-0002-4909-5763]{Akihiko Fukui}
\affiliation{Komaba Institute for Science, The University of Tokyo, 3-8-1 Komaba, Meguro, Tokyo 153-8902, Japan}
\affiliation{Instituto de Astrofisica de Canarias (IAC), 38205 La Laguna, Tenerife, Spain}
\email{Akihiko Fukui <afukui@g.ecc.u-tokyo.ac.jp>}

\author[0000-0001-8638-0320]{Andrew~W.~Howard}
\affiliation{Department of Astronomy, California Institute of Technology, Pasadena, CA 91125, USA}
\email{ahoward@caltech.edu}
\author[0000-0002-0531-1073]{Howard~Isaacson}
\affiliation{Department of Astronomy,  University of California Berkeley, Berkeley, CA 94720, USA}
\email{hisaacson@berkeley.edu}
\author[0000-0003-3504-5316]{Benjamin~J.~Fulton}
\affiliation{NASA Exoplanet Science Institute / Caltech-IPAC, Pasadena, CA 91125, USA}
\email{bjfulton@ipac.caltech.edu}
\author[0000-0002-8958-0683]{Fei~Dai}
\affiliation{Institute for Astronomy, University of Hawai‘i, 2680 Woodlawn Drive, Honolulu, HI 96822, USA}
\email{fdai@hawaii.edu}

\begin{abstract}

The ``Neptunian ridge'' is a recently identified peak in the frequency of planets with sizes between that of Neptune and Saturn orbiting their host stars with periods between 3 and 6 days \citep{Castro-Gonzalez2024}. These planets may have formed similarly to their larger, hot Jupiter counterparts in the ``three-day pile-up'', through a dynamically excited migration pathway.
The distribution of stellar obliquities in hot Neptune systems may therefore provide a vital clue as to their origin.
We report a new stellar obliquity measurement for TOI-2374\,b, a planet in the Neptunian ridge ($P = 4.31$~days, $\Rp = 7.5\,\Rearth$).
We observed a spectroscopic transit of TOI-2374\,b with the Keck Planet Finder, detecting the Rossiter-McLaughlin (RM) anomaly with an amplitude of 3~m/s, and measured a sky-projected obliquity of $\lambda = {\FitLambda}$, indicating an orbit significantly misaligned with the spin axis of its host star.
A reloaded RM analysis of the cross-correlation functions confirms this misalignment, measuring $\lambda = \FitLambdaRMReloaded$.
Additionally, we measured a stellar rotation period of $\Prot = \FitProt$~days with photometry from the \Tierras observatory, allowing us to deduce the three-dimensional stellar obliquity of $\psi = {\FitPsiRMReloaded}$.
TOI-2374\,b joins a growing number of hot Neptunes on polar orbits. The high frequency of misaligned orbits for Neptunian ridge and desert planets, compared with their longer period counterparts, is reminiscent of patterns seen for the giant planets and may suggest a similar formation mechanism.

\end{abstract}



\section{Introduction} 

The ``hot Neptune desert'' --- the paucity of planets with sizes between that of Saturn and Neptune at short orbital periods --- is one of the most striking features in the exoplanet landscape \citep{Szabo2011,Beauge2013,Mazeh2016}.
Intense mass loss due to photoevaporation is thought to clear out part of the hot Neptune desert and shape its lower edge \citep{Owen2013,Lopez2013}; however, the more massive planets along the upper edge of the hot Neptune desert are able to hold on to their gaseous atmospheres and do not lose significant mass fractions through photoevaporation \citep{Vissapragada2022a}.
Instead, the upper edge of the hot Neptune desert may be shaped by high-eccentricity migration and tidal disruption of planets that overflow their Roche lobe during pericenter approach \citep{Owen2018}.

Based on a reanalysis of data from the \Kepler mission, \citep{Castro-Gonzalez2024} recently identified a peak in planet occurrence at 3-6 days in orbital period, which they termed the ``Neptunian ridge'', separating the hot Neptune desert from longer-period planets of the same size.
This peak is similar to the ``three-day pile-up'' seen for hot Jupiters and may indicate a common, high-eccentricity migration pathway for these two types of planets.
A dynamically excited origin for these planets may also leave its imprint on the distribution of their stellar obliquities, and indeed, many hot Neptunes and sub-Saturns have been found to be on orbits significantly misaligned with their host stars' spin axes \citep[e.g.,][]{EspinozaRetamal2024}, although the sample of such measurements is still small.

NASA's TESS mission \citep{TESS_Ricker2014} is discovering thousands of transiting planets orbiting bright stars, even in relatively sparsely populated regions of the exoplanet census like the Neptunian ridge and desert.
TOI-2374\,b is one such newly-discovered sub-Saturn planet in the Neptunian ridge, with an orbital period of $P = 4.31$~days, radius $\Rp = 7.5\pm0.2\,\Rjup$, and mass $\Mp = 0.19\pm0.01\,\Mjup$ \citep{Hacker2024}.
In this manuscript, we report a new measurement of the three-dimensional stellar obliquity in this planetary system, using spectroscopic observations from the Keck Planet Finder and a measurement of the photometric rotation period from the \Tierras observatory, as part of a broader effort to measure three-dimensional obliquities in systems containing sub-Saturn planets \citep{Yee2025,Tamburo2025}.
We describe these observations in Section \ref{sec:observations}.
We perform a global characterization of the planetary system using  \TESS and ground-based light curves, out-of-transit radial velocity measurements, and broadband photometry, which is described in Section \ref{ssec:global_modeling}.
We analyze the in-transit spectroscopic observations using the canonical Rossiter-McLaughlin effect (\S\ref{ssec:rm_analysis}) as well as the reloaded RM effect (\S\ref{ssec:reloaded_rm}).
Finally, we discuss TOI-2374\,b in context of other sub-Saturn planets in Section \ref{sec:discussion}.

\section{Observations \label{sec:observations}}

\subsection{KPF Spectroscopy}

\begin{deluxetable}{ccc} \label{tab:rv_obs}
\tablecaption{KPF Radial Velocities for TOI-2374.}
\tablehead{
\colhead{Obs. Time\tablenotemark{a}} & \colhead{RV} & \colhead{$\sigma_\mathrm{RV}$} \\
\colhead{\bjdtdb} & \colhead{[\ms]} & \colhead{[\ms]}
}
\startdata
2460254.691822 & 15036.37 & 2.06 \\
2460254.773565 & 15029.79 & 1.73 \\
2460254.751810 & 15029.69 & 1.69 \\
2460254.811073 & 15025.61 & 2.15 \\
2460254.778927 & 15029.64 & 1.78 \\
2460254.681229 & 15033.82 & 2.30 \\
2460254.735509 & 15030.82 & 1.68 \\
2460254.784259 & 15030.02 & 1.81 \\
2460254.730120 & 15029.71 & 1.66 \\
2460254.714071 & 15037.21 & 1.85 \\
2460254.757349 & 15029.13 & 1.73 \\
2460254.849620 & 15022.30 & 2.73 \\
2460254.724683 & 15031.96 & 1.88 \\
2460254.854717 & 15022.08 & 2.98 \\
2460254.789727 & 15030.71 & 1.88 \\
2460254.746458 & 15027.43 & 1.65 \\
2460254.719384 & 15033.12 & 1.82 \\
2460254.833265 & 15029.81 & 2.34 \\
2460254.697820 & 15037.00 & 2.07 \\
2460254.740980 & 15030.89 & 1.69 \\
2460254.822390 & 15026.26 & 2.78 \\
2460254.795164 & 15029.76 & 1.86 \\
2460254.708295 & 15033.46 & 1.84 \\
2460254.686756 & 15033.59 & 1.95 \\
2460254.816665 & 15030.90 & 2.17 \\
2460254.828024 & 15027.49 & 2.62 \\
2460254.838493 & 15031.92 & 2.39 \\
2460254.806053 & 15029.65 & 1.93 \\
2460254.762791 & 15027.44 & 1.77 \\
2460254.703103 & 15036.43 & 1.85 \\
2460254.843950 & 15027.19 & 2.67 \\
2460254.768122 & 15029.15 & 1.69 \\
2460254.800783 & 15029.59 & 1.85 \\

\enddata
\tablenotetext{a}{Flux-weighted observation midpoint.}
\end{deluxetable}

We observed TOI-2374 with the Keck Planet Finder (KPF; \citealt{KPF_Gibson2020,KPF_Gibson2024}) on the night of UT 2023 Nov 06, during a transit of the planet TOI-2374\,b.
KPF is a stabilized, high-resolution (R $= 100{,}000$) fiber-fed spectrograph on the Keck I telescope on Mauna Kea. 
We observed in the standard instrument configuration beginning at astronomical twilight and continuing for four hours until the target set.
A total of 33 exposures were obtained with 420s exposure times each, with 13 of those occurring during the planet's transit.
We bracketed the observation sequence with calibration frames of a laser frequency comb and a ThAr lamp to improve the wavelength solution.

The data were reduced and 1D spectra were extracted using the standard KPF Data Reduction Pipeline (KPF-DRP; \citealt{KPF_Gibson2020}).\footnote{\url{https://github.com/Keck-DataReductionPipelines/KPF-Pipeline}}
The spectra were correlated against a weighted line mask specialized for stars with K2 spectral type \citep{Baranne1996,Pepe2002} to produce cross-correlation functions (CCFs) on an order-by-order basis, across both the green and red CCDs of KPF.
A combined CCF was computed by summing the CCFs weighted by the expected stellar flux in each echelle order.
Finally, the radial velocity (RV) and corresponding uncertainty were derived by fitting a Gaussian to the total CCF for each observation.
The RV measurements are provided in Table \ref{tab:rv_obs}.

\subsection{MuSCAT3 Transit Photometry}

We additionally obtained simultaneous photometry of TOI-2374 on 2023 Nov 06 to refine the mid-transit time of the planet.
We observed the star using the MuSCAT3 multicolor imager \citep{Muscat3_Narita2020} on the 2-m
Faulkes Telescope North at Haleakala Observatory in Hawaii, operated as part of the Las Cumbres Observatory Global Telescope network (LCOGT; \citealt{LCOGT_Brown2013}).
We observed simultaneously in the Sloan -$g$, -$r$, -$i$, and $z_s$ bands using exposure times of 16, 10, 10, and 15 seconds, respectively.
The data were reduced using the LCOGT {\tt BANZAI} pipeline \citep{McCully_2018SPIE10707E}  and aperture photometry performed using \texttt{AstroImageJ} \citep{AstroImageJ_Collins17}.
We also recorded auxiliary data including the width of the point-spread function, as well as the total flux in the comparison stars, that were used to detrend the light curves simultaneously with the light curve fits described in Section \ref{ssec:global_modeling}.
The planet's transits were confidently detected with consistent transit depths and shapes in all four bands (Figure \ref{fig:rm_plot}).

\subsection{\Tierras Photometry}

We monitored the brightness of TOI-2374 from the \Tierras observatory \citep{Tierras_GarciaMejia2020} from UT~5~October~2024 to \tierrasenddate, with the goal of measuring the stellar rotation period from flux modulations caused by the rotation of features on the stellar surface in and out of view.
We observed TOI-2374 as part of a broader program to measure rotation periods of stars with projected obliquity measurements in order to determine their true, three-dimensional obliquities \citep{Tamburo2025}.

All observations were performed with a 30-s exposure time in the custom $Tierras$ filter, a narrow-band filter in the near-infrared \citep[$\lambda_{\mathrm{C}}=863.5$~nm, $40.2$~nm FWHM;][]{Tierras_GarciaMejia2020} that was designed to limit photometric errors due to precipitable water vapor variability to less than $250$ parts-per-million (ppm). Each night, we conducted between one and five visits to the target of five minutes each, depending on observability and scheduling constraints. Data were reduced and photometry was performed using the custom $Tierras$ pipeline \citep{Tamburo2025}. We measured fluxes for all 210 sources in the images with Gaia $RP$ magnitudes less than 17 using circular apertures with radii of 5\textendash 20 pixels. We measured the local background for each source using the mean of the pixels within an annulus centered on the source with an inner radius of 35 pixels and and outer radius of 55 pixels, after applying a 2$\sigma$ clipping. We then generated relative photometry for each source using an iterative approach to determine optimal reference star weights for each source's artificial light curve \citep[ALC;][]{Broeg2005, Murray2020, Tamburo2022a, Tamburo2025}. The aperture size that minimized the scatter was chosen as the best light curve for each source. For TOI-2374, this was an aperture with a radius of nine pixels.

\subsection{Stellar Companion}
\begin{deluxetable}{lcc}
\tablecaption{Photometry and astrometry of TOI-2374 and its companion from \textit{Gaia} DR3 \label{tab:gaia_comp_props}}
\addtocounter{table}{1}
\pdfbookmark[3]{Table \thetable: Stellar Companions}{gaia_comp_props}%
\addtocounter{table}{-1}
\tabletypesize{\footnotesize}
\tablehead{
    & \colhead{Primary} & \colhead{Secondary} \\
    & \colhead{HD 202673} & \colhead{TOI-2374}
}
\startdata
Gaia DR3 ID & 6828814283414902784 & 6828814283414902912 \\
TIC ID & 439366537 & 439366538 \\
Ang. Sep. ($"$) & 22.34 & -- \\
Proj. Sep. (AU) & 3040 & --\\
Parallax (mas) & $7.372 \pm 0.019$ & $7.394 \pm 0.018$ \\
$\mu_{{\alpha}}$ (mas/yr) & $-11.949 \pm 0.018$ & $-11.470 \pm 0.018$ \\
$\mu_{{\delta}}$ (mas/yr) & $-31.746 \pm 0.018$ & $-31.421 \pm 0.016$ \\
RV (km/s) & $14.15 \pm 0.17$ & $14.77 \pm 0.29$ \\
$G$ (mag) & $9.223 \pm 0.003$ & $11.819 \pm 0.003$ \\
$G_\mathrm{BP}$ (mag) & $9.489 \pm 0.003$ & $12.346 \pm 0.003$ \\
$G_\mathrm{RP}$ (mag) & $8.795 \pm 0.004$ & $11.144 \pm 0.004$
\enddata
\end{deluxetable}
\citet{GaiaEDR3_Binaries_El-Badry2021} published a catalog of wide binary stars determined to be likely bound companions from their \textit{Gaia} astrometry and radial velocities.
The planet host star TOI-2374 appears in this catalog --- the nearby star HD~202673 (Gaia DR3 6828814283414902784) has nearly identical parallax, proper motion, and radial velocity (Table \ref{tab:gaia_comp_props}).
\citet{GaiaEDR3_Binaries_El-Badry2021} computed an approximate chance alignment probability for the pair of $\mathcal{R} = 3\times10^{-5}$, so this is almost certainly a bound pair of stars.
At their measured distance, the two stars' sky separation of 22$^{\prime\prime}$ corresponds to a projected separation of 3000 AU.
Interestingly, the companion to TOI-2374 is a more massive F9 star, making this one of only a handful of binary systems where the secondary component is known to host a transiting planet.
In the rest of the manuscript, we refer to the planet host as TOI-2374B or TOI-2374 for consistency with the previous literature, the planet as TOI-2374\,b or TOI-2374B\,b, and the primary star in the wide binary as TOI-2374A or HD~202673.

\subsection{Archival Data}

In addition to the new observations discussed above, we incorporated the data previously published by \citet{Hacker2024} in our analysis.
We used photometric measurements from TESS, which observed TOI-2374 in Sectors 1 and 28 at 1800s and 120s cadence respectively. Specifically, we made use of the light curve produced by the Quick Look Pipeline (QLP; \citealt{TESS_QLP_Huang2020a,TESS_QLP_Huang2020b}) for the Sector 1 observations, and that produced by the TESS Science Processing Operations Center (SPOC; \citealt{TESS_SPOC_Jenkins2016}) for Sector 28.
\citet{Hacker2024} also published ground-based follow-up photometry from the Brierfield Observatory in New South Wales, Australia, as well as the South Africa Astronomical Observatory (SAAO) and Siding Springs Observatory (SSO) nodes of the LCOGT network.
To measure the planet's mass, \citet{Hacker2024} used precise RV measurements from the High Accuracy Radial velocity Planet Searcher \citep[HARPS;][]{HARPS_Mayor2003} and Planet Finder Spectrograph \citep[PFS;][]{PFS_Crane2006,PFS_Crane2008,PFS_Crane2010}, and we incorporated the published RV measurements in our analysis.

\section{Analysis}

We describe our analysis of the TOI-2374 system in the following section.
First, we measured the stellar rotation period using the \Tierras photometry (\S\ref{ssec:tierras_rotation}).
We then modeled the TESS and ground-based photometry, broadband fluxes, and out-of-transit RVs, 
to improve the characterization of the stellar properties and planet's orbital parameters, primarily the transit ephemeris at the epoch of the KPF observations (\S\ref{ssec:global_modeling}).
The results from those two analyses were used as priors for our measurement of the 3D stellar obliquity of TOI-2374, which we derived both using the canonical Rossiter-McLaughlin (RM) effect (\S\ref{ssec:rm_analysis}) and with a ``reloaded'' RM \citep{Cegla2016} analysis (\S\ref{ssec:reloaded_rm}), finding TOI-2374\,b to be on a polar orbit around its host star.

\subsection{Rotation Period Measurement} \label{ssec:tierras_rotation}

We measured a rotation period from the \Tierras data following procedures similar to those described in \citet{Tamburo2025}. We first applied quality cuts to the data, retaining exposures with FWHM seeing values less than $4\arcsec$, pointing errors less than 20 pixels along both the R.A. and decl. axes, and world coordinate system solutions with RMS $< 0.215\arcsec$ (half a \textit{Tierras} pixel). We also applied a cut on the median ALC flux of each image. We measured a significant decrease (about 35\%) in the median flux of sources across the $Tierras$ time series due to the dust accumulation on the mirror. We identified points corresponding the upper envelope of the median ALC time series using the \texttt{argrelextrema} function from \texttt{scipy} \citep{Scipy} with a window of 20 points, then fit a linear function to these upper envelope points to correct for the loss of flux due to the aforementioned dust accumulation. We retained exposures where the corrected median ALC had a flux greater than 0.9. With these quality cuts applied, the $Tierras$ light curve consisted of $753$ exposures across $63$ individual nights. 

\begin{figure*}
    \centering
    \includegraphics[width=\textwidth]{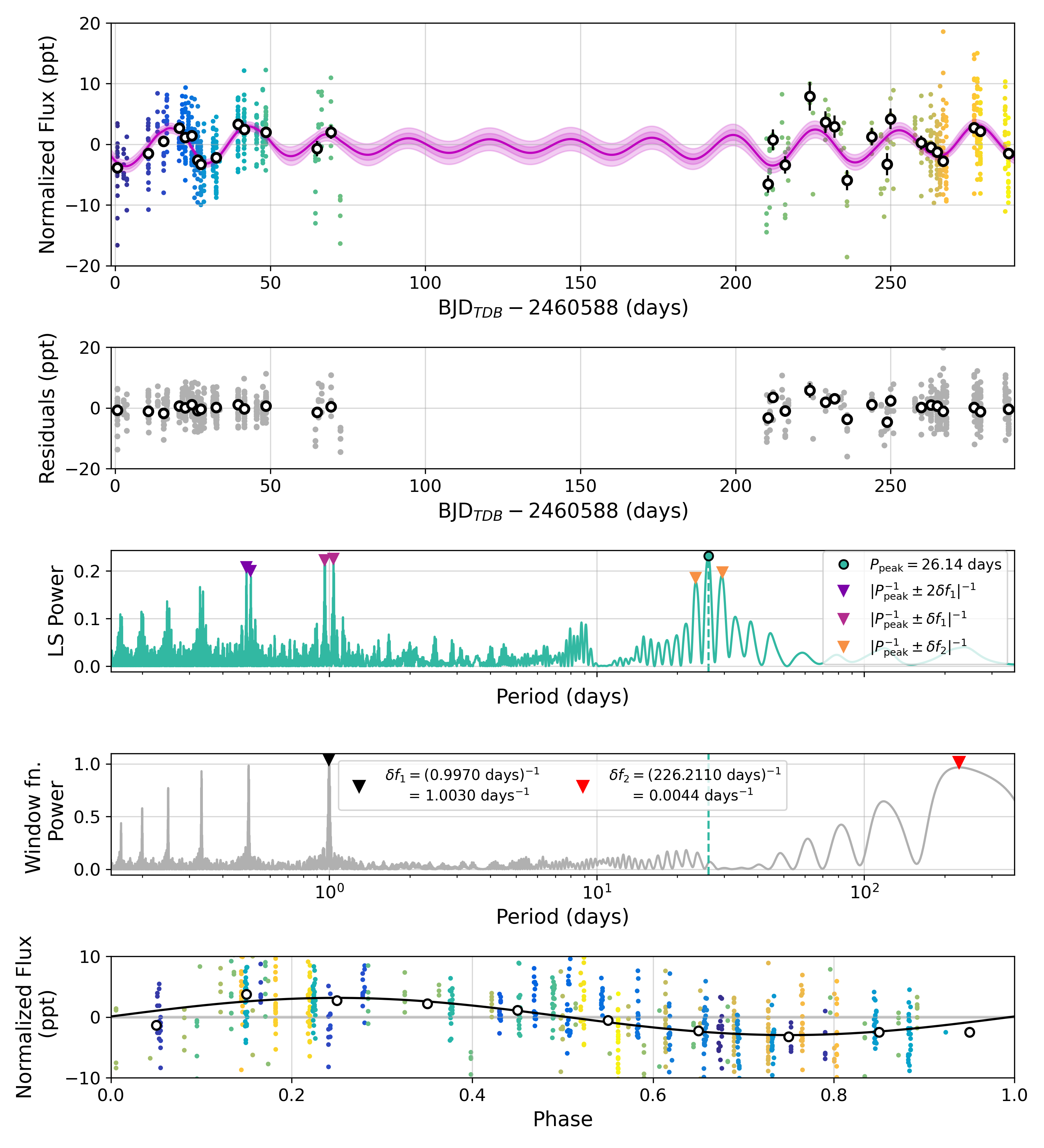}
    \caption{First row: The $Tierras$ light curve at the native 30-s cadence (great points) along with the best-fit QP-GP model (magenta). The data are colored by time and have been corrected with a linear function of sky background given by $c_0+c_1\cdot \mathrm{sky}$. The 1- and 2-$\sigma$ uncertainty regions of the best-fit QP-GP model are shown as shaded regions. The white points show the data binned over two-night intervals. Second row: The residuals from the best-fit QP-GP model. The residual scatter is larger between days 210 and 245 due to a smaller number of visits per night during that time period. Third row: The LS periodogram of the data. The peak of the periodogram is at $P_\mathrm{peak} = 26.14$~days. The next six highest peaks are indicated with triangles. These peaks are  expected aliases of $P_\mathrm{peak}$ given the periods at which the window function (fourth row) has significant power. Fifth row: The data phase-folded on a period of $26.38$~days, the rotation period inferred from the GP model. The color of the points matches those in the first row and indicate the progression of time. White points show the data binned over equal widths in phase. We plot a sine model in black to show the periodic nature of the variability.}
    \label{fig:gp}
\end{figure*}

We show the data in the top row of  Figure~\ref{fig:gp}. We performed a Lomb-Scargle (LS) periodogram of the data using the implementation in \texttt{astropy} \citep{Astropy13, Astropy18, Astropy2022}, which is plotted in the third row of Figure~\ref{fig:gp}. This revealed a significant peak at $P_\mathrm{peak} = 26.14$~days. Following \citet{Vanderplas2018}, we computed the window function of the data, which had significant power at $df_1^{-1}=0.9969$~days and $df_2^{-1}=222.6293$~days (the peaks in the window function with periods less than 1 day are aliases of the peak at 0.9969~days). We then computed the expected aliases of the 26.14-day LS periodogram peak given the window function peaks $df_1$ and $df_2$. As the third row of Figure~\ref{fig:gp} shows, the remaining significant peaks can be identified with these aliases. Folding the data on $P_\mathrm{peak}$ reveals clear sinusoidal variation (see the last row of Figure~\ref{fig:gp}). In combination with the alias analysis, we therefore conclude that the rotation period of TOI-2374 is indeed about 26~days.
We also performed the same analysis on the two seasons of \textit{Tierras} data separately and detected a $\sim 26$ day signal in each season's data, albeit at a lower power than in the combined dataset.


To estimate the uncertainty on the rotation period, we performed a Gaussian process (GP) analysis similar to that described in \citet{Tamburo2025}. We used the \texttt{python} package \texttt{celerite} \citep{ForemanMackey2017} to generate GP models with a quasi-periodic (QP) covariance function of the form

\begin{equation}
    \kappa(\tau) = \frac{B}{2+C}\exp^{-\tau/L}\left [ \cos{\frac{2\pi \tau}{P_\mathrm{rot}} + (1 +C)}\right ].
\end{equation}

Here, $\tau$ is an array of absolute differences between timestamps, with $\tau_{\mathrm{ij}}=|x_\mathrm{i} - x_\mathrm{j}|$. $B$ and $C$ control the amplitude of the covariance, $L$ represents  the exponential decay timescale, and $P_\mathrm{rot}$ is the rotation period of TOI-2374. It has been demonstrated that the QP kernel can serve as a reliable estimator of the rotation periods of stars \citep[e.g., ][]{Angus2018, Nicholson2022}.

We sampled different values for $\ln B$, $\ln C$, $\ln L$, and $\ln P_\mathrm{rot}$ conditioned on our data using \texttt{emcee} \citep{emcee_ForemanMackey2013}. We placed uniform priors on the parameters as given in Table~\ref{tab:gp}. We also added a function of sky background to the computed QP-GP flux of the form $c_0 + c_1\cdot \mathrm{sky}$ \citep{Tamburo2025}. Finally, we fit for a factor $f$ which represents the factor by which the pipeline errors were under- or overestimated. We ran 100 walkers and terminated the sampling when the number of steps was greater than 100$\times$ the autocorrelation time, which occurred at $15,800$ steps. We discarded three times the average autocorrelation time to account for burn-in and thinned the chains by half the mean autocorrelation time. In total, we were left with $23,700$ samples for each parameter. 

\begin{deluxetable}{llr} \label{tab:gp}
\tablecaption{Priors and posteriors for parameters in our QP-GP analysis}
\tablehead{\colhead{Parameter} & \colhead{Prior} & 
\colhead{Posterior}}
\startdata
$\ln(B/\mathrm{ppt}^2)$ & $\mathcal{U}(-20.0,0.0)$ & $-0.052^{+0.039}_{-0.085}$\\
$\ln(C)$ & $\mathcal{U}(-20.0,20.0)$ & $-10.8\pm 6.3$ \\
$\ln(L/\mathrm{day})$ & $\mathcal{U}(4.0, 10.0)$ & $4.13^{+0.10}_{-0.22}$\\
$P_\mathrm{rot}$ (days) & $\mathcal{U}( 20.0,  35.0)$ & $26.38^{+0.90}_{-0.80}$\\
$f$ & $\mathcal{U}(0, \infty)$ & $1.39^{+0.038}_{-0.037}$\\
$c_0$ & $\mathcal{U}(-\infty, \infty)$ & $1.30^{+0.37}_{-0.36}$\\
$c_1$ & $\mathcal{U}(-\infty, \infty)$ & $-0.57\pm0.14$ \\
\enddata
\end{deluxetable}

We report the median value for the parameters in Table~\ref{tab:gp} along with uncertainties as given by the 16\textendash 84 percentile range of each posterior distribution. We measure a rotation period of $P_\mathrm{rot}={\FitProt}$~days, consistent with the value of $P_\mathrm{peak} = 26.14$~days determined through the LS periodogram analysis. In the remainder of our analysis, we adopt the value of $P_\mathrm{rot}$ from the QP-GP analysis and its corresponding uncertainty as the rotation period for TOI-2374. We show the best-fit QP-GP model in the top panel of Figure~\ref{fig:gp}, with the residuals after subtracting the GP model shown in the second panel. The unbinned residuals of the best-fit light curve have a standard deviation of $4.3$~parts-per-thousand (ppt), while the residuals binned over two-night intervals have a standard deviation of $2.0$~ppt. 

We also checked if the photometric rotation signal detected by \textit{Tierras} was also detected in the TESS data. We computed Lomb-Scargle periodograms for the two TESS sectors individually as well as a combined light curve after masking out the planet's transits, but found no significant peaks. Given that our inferred rotation period of $\Prot = 26.4$ days is comparable to the duration of each TESS sector, and that TOI-2374 was only observed by TESS in two individual sectors spaced by two years, it is unsurprising that a periodic rotation signal could not be detected in the TESS data.
Previous analyses of TESS data show that it is challenging to measure stellar rotation periods from TESS data when $\Prot$ is greater than half a TESS sector, except in cases where the star is observed in multiple consecutive sectors \citep[e.g.,][]{Claytor2024,Hattori2025}.
Indeed, our observing program to measure stellar rotation periods with the \textit{Tierras} observatory is focused on sub-solar-mass stars whose slow rotation would not be easily detected by TESS.

We note that \citet{Hacker2024} reported a rotation period of $50.2\pm2.8$~days for TOI-2374, which is approximately a factor of two times our measurement ($P_\mathrm{Hacker}/P_\mathrm{rot} = 1.90\pm 0.13$).
However, this was derived from the relationship between the activity indicator \logrhk and stellar rotation calibrated by \citet{Noyes1984}, and the uncertainty did not account for the considerable scatter in this relationship. 
Furthermore, the measured $\logrhk=-5.0$ measured for TOI-2374 is on the edge of the calibration sample used by \citet{Noyes1984}, potentially affecting the reliability of the calibration for stars as inactive as TOI-2374.

Because stars spin down with time due to magnetic braking, gyrochronology allows us to infer a star's age given its rotation period and mass.
We used the \texttt{gyro-interp} code \citep{Gyrointerp_Bouma2023}, which interpolates between empirically measured rotation period sequences from stellar clusters with known ages, to derive a stellar age of $3.1 \pm 0.3$~Gyr for TOI-2374.
Specifically, we used v0.6 of \texttt{gyro-interp}, which extends the calibration out to 4~Gyr using the rotation sequence from M67 \citep{Gyrointerp_Bouma2024}, and incorporates a statistical age floor at older ages to account for increasing scatter in the rotation period distribution.
This age derived from gyrochronology is consistent with the age of \FitAge derived by fitting the MIST isochrones to the broadband photometry (\S\ref{ssec:global_modeling}).
While the uncertainties derived by \texttt{gyro-interp} include the statistical uncertainty arising from the intrinsic scatter in rotation periods as measured from cluster measurements, they do not account for systematic biases.
The most notable of these is the potential tidal spin up of hot Jupiter host stars by their planets \citep{Brown2014,Maxted2015}, although TOI-2374\,b may have too low a mass to significantly spin up its host star \citep[e.g.,][]{TejadaArevalo2021}.

\subsection{Global System Modeling} \label{ssec:global_modeling}

\begin{deluxetable*}{lccc} \label{tab:exofast_results}
\tablecaption{Median values and 68\% Confidence Intervals from \Exofast fit of TOI-2374}
\tablehead{\colhead{Parameter} & \colhead{Description} & \multicolumn{2}{c}{Value}}
\startdata
\\[-\normalbaselineskip]\multicolumn{2}{l}{Stellar Parameters:} & TOI-2374 & HD 202673 \\
~~~~$M_\star$ ($M_\odot$) & Stellar mass & $0.802^{+0.029}_{-0.028}$ & $1.183^{+0.063}_{-0.079}$ \\
~~~~$R_\star$ ($R_\odot$) & Stellar radius & $0.711 \pm 0.015$ & $1.465^{+0.067}_{-0.063}$ \\
~~~~$\log{g_\star}$ (cgs) & Stellar surface gravity & $4.638 \pm 0.017$ & $4.177^{+0.048}_{-0.051}$ \\
~~~~$\rho_\star$ (g cm$^{-3}$) & Stellar density & $3.14^{+0.18}_{-0.17}$ & $0.527^{+0.087}_{-0.076}$ \\
~~~~$L_\star$ ($L_\odot$) & Stellar luminosity & $0.285^{+0.013}_{-0.012}$ & $2.67^{+0.15}_{-0.13}$ \\
~~~~$T_\mathrm{eff}$ (K) & Stellar effective temperature & $5002^{+65}_{-64}$ & $6100 \pm 140$ \\
~~~~$[\mathrm{Fe/H}]$ (dex) & Metallicity & $0.128^{+0.082}_{-0.076}$ & $0.026^{+0.10}_{-0.094}$ \\
~~~~$[\mathrm{Fe/H}]_0$ (dex)\tablenotemark{a} & Initial metallicity & $0.105^{+0.078}_{-0.071}$ & $0.105^{+0.078}_{-0.071}$ \\
~~~~Age (Gyr)\tablenotemark{a} & Stellar age & $4.5^{+2.1}_{-1.4}$ & $4.5^{+2.1}_{-1.4}$ \\
~~~~EEP & Equal evolutionary phase & $327^{+10}_{-11}$ & $408^{+28}_{-33}$ \\
~~~~$A_V$ (mag)\tablenotemark{a} & Visual extinction & $0.065 \pm 0.044$ & $0.065 \pm 0.044$ \\
~~~~d (pc)\tablenotemark{a} & Distance & $135.22 \pm 0.33$ & $135.22 \pm 0.33$ \\
\\[-\normalbaselineskip]\multicolumn{2}{l}{Planet Parameters:} & TOI-2374\,b \\
~~~~$P$ (days) & Period & $4.3136193 \pm 0.0000015$ \\
~~~~$T_c$ (BJD$_\mathrm{TDB}$) & Time of conjunction & $2460069.26439 \pm 0.00013$ \\
~~~~$R_P$ ($R_\mathrm{J}$) & Planet radius & $0.668 \pm 0.018$ \\
~~~~$M_P$ ($M_\mathrm{J}$) & Planet mass & $0.194^{+0.014}_{-0.015}$ \\
~~~~$\left(R_P / R_\star\right)^2$ & Planet-star area ratio & $0.00933 \pm 0.00015$ \\
~~~~$K$ (m/s) & RV semi-amplitude & $28.0^{+1.9}_{-2.0}$ \\
~~~~$a$ (AU) & Semimajor axis & $0.04818^{+0.00057}_{-0.00058}$ \\
~~~~$a/R_\star$ & Planet-star separation & $14.57^{+0.27}_{-0.26}$ \\
~~~~$i$ (deg) & Inclination & $87.20 \pm 0.10$ \\
~~~~$b \equiv a\cos{i}/R_\star$ & Transit impact parameter & $0.711^{+0.013}_{-0.014}$ \\
~~~~$\rho_P$ (g cm$^{-3}$) & Planet density & $0.803^{+0.086}_{-0.080}$ \\
~~~~$\log{g_P}$ (cgs) & Planet surface gravity & $3.031^{+0.036}_{-0.039}$ \\
~~~~$T_\mathrm{eq}$ (K) & Planet equilibrium temperature & $926 \pm 10.000000$ \\
~~~~$\langle F \rangle$ (10$^9$ erg s$^{-1}$ cm$^{-2}$) & Incident flux & $0.1672^{+0.0075}_{-0.0072}$ \\
~~~~$T_{14}$ (days) & Transit duration & $0.07885 \pm 0.00060$ \\
~~~~$\tau$ (days) & Ingress/egress duration & $0.01310^{+0.00059}_{-0.00057}$%
\enddata

\tablenotetext{a}{These parameters are linked for both stellar components.}
\end{deluxetable*}

Modeling the Rossiter-McLaughlin effect requires precise knowledge of when the planet's transit occurs.
Because of the significant time gap between the last photometric observations obtained by \citet{Hacker2024} and our KPF observations, the transit ephemeris published by \citet{Hacker2024} propagated to the epoch of the KPF observations was uncertain to $\pm9$~minutes.
We therefore reanalyzed the previously published TESS and ground-based photometry together with our new MuSCAT3 photometry, extending the total observing baseline to five years.

We used the \Exofast code \citep{ExoFASTv2_Eastman19} to perform this reanalysis. \Exofast self-consistently models the stellar and planetary properties, taking advantage of the transit-based constraint on the stellar density to significantly improve the constraints on the stellar radius and mass \citep[e.g.,][]{Eastman2023}. In addition to the transit photometry, we also included as input to the global analysis the previously-published HARPS and PFS RVs, along with broadband flux measurements from \Gaia DR3 \citep{GaiaEDR3_Riello2021}, 2MASS \citep{TMASS_Cutri2003}, WISE \citep{WISE_Cutri2012}, and APASS \citep{APASS_2016yCat.2336....0H}, the latter of which are used to constrain the stellar spectral energy distribution.

The overall fitting strategy we used is similar to that described in \citet{Yee2025}.
We fixed the planet's orbit to be circular given its planet's short orbital period and that the previous analysis found no evidence for any significant eccentricity.
We simultaneously modeled both components of the TOI-2374AB system under the assumption that the two stars are a bound, coeval pair with the same age and initial metallicity.
We placed a Gaussian prior on the current metallicity of TOI-2374 of $\feh = +0.15\pm0.04$~dex as reported by \citet{Hacker2024} from their analysis of the HARPS spectra.
By simultaneously modeling both TOI-2374 and its more massive, faster evolving companion, we are able to obtain a better age constraint on the stellar system compared with modeling a single star, as done by \citet{Hacker2024}.
We used uniform priors on the remaining fitting parameters.
We found best-fit parameters and estimated the posterior probability distributions for each parameter using a differential evolution Monte Carlo Markov Chain (MCMC) procedure.

We report the median values and 68\% confidence intervals from our global \Exofast fit in Table \ref{tab:exofast_results}.
Crucially, 
we achieved a mid-transit timing precision of just 13~seconds at the epoch of the KPF RV observations, a $>40\times$ improvement over the ephemeris published by \citet{Hacker2024}.
We also constrain the stellar radius and mass to a precision of 2\% and 4\% respectively.

\subsection{Rossiter-McLaughlin Analysis} \label{ssec:rm_analysis}

\begin{figure}
\epsscale{1.2}
\plotone{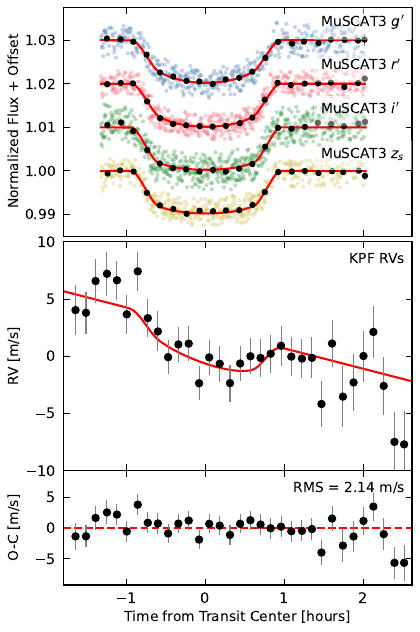}
\caption{KPF RVs showing the detection of the Rossiter-McLaughlin effect for TOI-2374\,b (bottom panel), along with contemporaneous multi-color photometry from MuSCAT3 (top panel).
Red lines show the maximum-likelihood model from the respective \Exofast and \rmfit fits.
The KPF RV data show a clear anomaly during the time of the photometric transit.
\label{fig:rm_plot}}
\end{figure}

We first measured the stellar obliquity with respect to TOI-2374\,b's orbit by modeling the KPF RVs with the canonical Rossiter-McLaughlin (RM) effect, using the \rmfit code \citep{rmfit_Stefansson2022}.
The RM effect describes the RV anomaly during a planet's transit as it occults parts of the stellar surface that have been blue- or red-shifted due to stellar rotation, and its shape and amplitude depend primarily on the sky-projected stellar obliquity $\lambda$ and the projected stellar rotation velocity \veqsini.
Instead of fitting for \veqsini, we parameterize the equatorial rotation velocity in terms of the stellar radius and rotation period $\veq = 2\pi \Rstar/\Prot$, where \Rstar is constrained from the \Exofast fit (\S\ref{ssec:global_modeling}) and \Prot from the \Tierras data (\S\ref{ssec:tierras_rotation}).
We allow the stellar inclination to vary, placing a uniform prior on $\cos{i_\star}$ as appropriate for an isotropic distribution of $i_\star$.
Our knowledge of the stellar rotation period enables us to fit for the stellar inclination, which we then use to compute the true, three-dimensional stellar obliquity:
\begin{equation} \label{eq:psi_eq}
    \cos\psi = \sin i_\star\sin i_\mathrm{orb} \cos\lambda + \cos i_\star \cos i_\mathrm{orb},
\end{equation}
where $i_\mathrm{orb}$ is the planet's orbital inclination as measured from the transit fit to the photometric data.
Modeling \veqsini in terms of \istar and \Prot ensures that the statistical dependence between \veq and \veqsini is accounted for when deriving the posterior probability distribution of $\psi$ \citep{Masuda2020}.

We compute the shape of the RM anomaly using the formulation described by \citet{Hirano2010}.
In addition to $\lambda$ and $\psi$, the \citet{Hirano2010} model includes an additional parameter, $\beta$, corresponding to the spectral line width as broadened by sources other than stellar rotation.
We place a Gaussian prior on $\beta$ centered at $3.1~\kms$ and with dispersion $1~\kms$, based on broadening due to the instrumental line profile and macroturbulence, where we computed the latter given the stellar \Teff of TOI-2374 using the relation from \citet{Valenti2005}.

For the additional parameters in the fit, such as the planet's RV semi-amplitude $K$, which governs the overall RV slope during the observation sequence, as well as the various orbit and transit parameters, we placed Gaussian priors based on the results derived in the global \Exofast modeling, with the following exceptions.
We used the exoCTK web tool\footnote{\url{https://exoctk.stsci.edu/limb_darkening}} to compute the limb-darkening coefficients and centered Gaussian priors with width 0.1 on those values.
Similar to \citet{Handley2025} and \citet{Tamburo2025}, we found that the instrumental uncertainties for the KPF RVs may have been overestimated.
To account for this, we fitted for a squared jitter term $\sigma_{\mathrm{J,KPF}}^2$, which we allowed to take on both positive and negative values. This term was then added to the instrumental uncertainties to obtain a modified total uncertainty, $\sigma_\mathrm{tot}^2 = \sigma_\mathrm{inst}^2+\sigma_{\mathrm{J,KPF}}^2$.
We found that $\sigma_\mathrm{tot}^2$ had to be reduced by $-0.5\,\mathrm{m}^2\,\mathrm{s}^{-2}$ to bring the reduced $\chi^2$ of the residuals to 1, corresponding to an $\approx10\%$ decrease in the per-point RV uncertainties.
We report the list of fit parameters and any priors imposed in Table \ref{tab:rm_results}.

We used the differential evolution \citep{PyDE_Storn1997} optimization code implemented in \texttt{PyDE}\footnote{\url{https://github.com/hpparvi/PyDE}} to find the maximum-likelihood RM model and corresponding parameters.
We derived uncertainties on the model parameters by sampling the posterior probability distributions using the \texttt{emcee} MCMC sampler \citep{emcee_ForemanMackey2013}.
The best-fit results and uncertainties are reported in Table \ref{tab:rm_results} and plotted in Figure \ref{fig:rm_plot}.

The KPF RV data show a clear deviation from the expected Keplerian motion of the star during the observed photometric transit of the planet. 
The best-fit RM model has an amplitude of $3.2\,\ms$, and we infer that TOI-2374\,b is on an orbit highly misaligned with the stellar spin axis, with a projected stellar obliquity of $\lambda = \FitLambda$.
The low amplitude of the RM anomaly, together with the rotation period measured from \Tierras, imply we are observing the star nearly pole-on. 
Indeed, the inferred stellar inclination is $\istar = {\FitIstar}$, with a projected rotation velocity of $\vsini = \FitVsini\,\kms$.
Combining the measurements of $\lambda$ and $\istar$ reveals that the true stellar obliquity is $\psi = {\FitPsi}$, and TOI-2374\,b belongs to the class of sub-Saturns on polar orbits.

\begin{deluxetable*}{lcc} \label{tab:rm_results}
\tablecaption{Priors and Fit Results for RM Analysis}
\tablehead{\colhead{Parameter} & \colhead{Prior} & \colhead{Posterior}}
\startdata
$T_c$ (BJD$_\mathrm{TDB}$) & $\mathcal{N}(2460254.74974,0.00015)$ & $2460254.74974 \pm 0.00015$ \\
$P_\mathrm{orb}$ (days) & $\mathcal{N}(4.3136193,0.0000015)$ & $4.3136193 \pm 0.0000015$ \\
$\lambda$ (deg) & $\mathcal{U}(-180,180)$ & $81^{+23}_{-22}$ \\
$\psi$ (deg) & Derived & $84.6^{+7.1}_{-7.3}$ \\
$P_\mathrm{rot}$ (days) & $\mathcal{N}(26.4,1.0)$ & $26.5 \pm 1.0$ \\
$\cos{i_\star}$ & $\mathcal{U}(0,1)$ & $0.943^{+0.024}_{-0.034}$ \\
$i_\star$ (deg) & Derived & $19.5^{+5.1}_{-4.7}$ \\
$v\sin{i_\star}$ (km/s) & Derived & $0.46 \pm 0.11$ \\
$b$ & $\mathcal{N}(0.714,0.013)$ & $0.716 \pm 0.013$ \\
$i_\mathrm{orb}$ (deg) & Derived & $87.175^{+0.071}_{-0.074}$ \\
$R_p/R_\star$ & $\mathcal{N}(0.09673,0.000765)$ & $0.09669^{+0.00076}_{-0.00075}$ \\
$R_\star$ ($R_\odot$) & $\mathcal{N}(0.717,0.015)$ & $0.717 \pm 0.015$ \\
$\beta$ (km/s) & $\mathcal{N}(3.1,1.0)$ & $3.06 \pm 0.99$ \\
$K$ (m/s) & $\mathcal{N}(27.8,2.1)$ & $29.6^{+1.9}_{-2.0}$ \\
$\gamma_\mathrm{KPF}$ (km/s) & $\mathcal{U}(-1000,1000)$ & $2.48^{+0.44}_{-0.46}$ \\
$\sigma^2_{\mathrm{KPF}}$ (m$^2$/s$^2$) & $\mathcal{U}(-10,10)$ & $-0.5^{+1.2}_{-0.8}$ \\
$u_{1,\mathrm{KPF}}$ & $\mathcal{N}(0.64,0.1)$ & $0.66 \pm 0.10$ \\
$u_{2,\mathrm{KPF}}$ & $\mathcal{N}(0.08,0.1)$ & $0.096 \pm 0.099$%
\enddata

\end{deluxetable*}

\subsection{Reloaded RM Analysis} \label{ssec:reloaded_rm}

We also analyzed the KPF data using the ``Reloaded Rossiter-McLaughlin'' technique \citep{Cegla2016}, which directly measures the stellar line profile occulted by the planet as the planet's shadow crosses the stellar surface.
Such an analysis makes use of the actual distortion to the spectral lines rather than the impact of this distortion on the lines' centroid positions as measured by shifts in the stellar RVs.

We made use of the cross-correlation functions (CCFs) generated by the KPF-DRP, normalized to the continuum for each observation.
We computed the systemic velocity from the centroid positions of the CCFs from the observations taken outside the transit.
We then shifted each CCF by the systemic velocity and the star's expected Keplerian motion such that each is centered at zero.
To place the CCFs on a consistent flux scale, we used the \texttt{batman} code \citep{Batman_Kreidberg15} to generate a transit light curve model using the best-fit transit parameters from \S\ref{ssec:global_modeling}, and multiplied the CCFs by the integrated stellar flux over each observation.
We constructed a template CCF ($\mathrm{CCF}_\mathrm{out}$) by summing the CCFs from the out-of-transit observations, representing the stellar line profile convolved with instrumental and atmospheric broadening effects.
We excluded the first two observations from the template as they were taken close to twilight and showed some evidence of solar contamination.
We constructed template CCFs for the green and red CCDs of KPF separately, and then combined the two by taking the weighted average according to their resepective uncertainties.

\begin{figure}
\epsscale{1.2}
\plotone{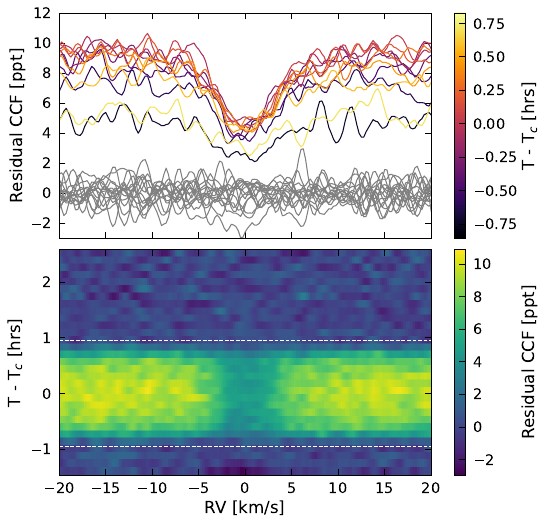}
\caption{\textbf{Top:} Residual CCFs obtained by subtracting each observation's CCF from the out-of-transit template. The in-transit CCFs are colored according to the time from mid-transit. For reference, the residual CCFs for the out-of-transit observations are shown in gray, showing no significant deviations above the noise level.
\textbf{Bottom:} This panel shows the residual CCFs with time flowing upwards on the vertical axis. Horizontal dashed lines show the planet ingress and egress times (first and fourth contacts).
\label{fig:rm_reloaded_ccfs}}
\end{figure}

\begin{figure}
\plotone{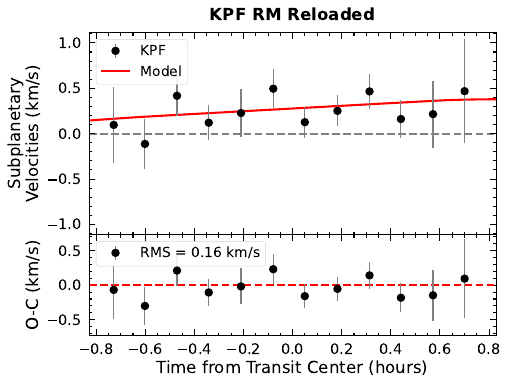}
\caption{Local RVs computed from the centroids of a Gaussian fit to the local CCFs.
The red line shows the best-fitting model to the local RVs, assuming solid body rotation for the star.
\label{fig:rm_reloaded}
}
\end{figure}

We isolated the ``local'' CCF by subtracting each in-transit CCF from the template CCF ($\mathrm{CCF}_\mathrm{loc} = \mathrm{CCF}_\mathrm{out} - \mathrm{CCF}_\mathrm{in}$).
This residual $\mathrm{CCF}_\mathrm{loc}$ corresponds to the stellar line profile arising from the region of the stellar surface behind the planet during its transit.
Figure \ref{fig:rm_reloaded_ccfs} shows these residual CCFs as a function of time from mid-transit.
We then fitted each $\mathrm{CCF}_\mathrm{loc}$ with a Gaussian profile to measure the subplanetary velocity $\mathrm{RV}_\mathrm{loc}$ (Figure \ref{fig:rm_reloaded}).
We derived $\mathrm{RV}_\mathrm{loc}$ and its corresponding uncertainty using the non-linear least squares \texttt{curve\_fit} algorithm from the \texttt{scipy} package \citep{Scipy}.
The local RVs show a consistent redshift, indicating that the planet only traverses the same hemisphere of the star and corroborating the measurement of the polar orbit from the canonical RM effect.

We modified \rmfit to model the local RVs by performing a flux-weighted integration of the stellar velocity field over the region occulted by the planet (Equation 9 of \citealt{Cegla2016}).
Similar to \citet{Rubenzahl2024a}, we performed the integration using a Fibonacci spiral grid, which allowed for faster convergence.
When fitting the reloaded RM model, we only included points during which the center of the planet's shadow overlapped the stellar disk for the entire exposure (corresponding to a positive limb angle $\mu$ for the full exposure).
In this analysis, the only free parameters were $\lambda$ and \veqsini, as the other planet and orbital parameters were fixed during the extraction of the subplanetary velocity.
Figure \ref{fig:rm_reloaded} shows the best-fit model assuming solid body rotation.

We also considered more complex models incorporating the effects of differential rotation and center-to-limb variations in the convective blueshift, but such models did not provide any appreciable improvement to the fit to justify the increase in the number of free parameters.
The lack of detection of center-to-limb variations is likely due to insufficient precision of the $\mathrm{RV}_\mathrm{loc}$ measurements and limited time sampling over the transit duration, especially while the planet's shadow occults the limbs of the star.
Detection of differential rotation as the planet traverses different stellar latitudes would in principle provide an constraint on the stellar inclination independent of the rotation period, but the low stellar inclination limits our ability to measure it in the case of TOI-2374.

The reloaded RM analysis yield results consistent with those obtained by the canonical RM method to within 1$\sigma$.
We find $\lambda = {\FitLambdaRMReloaded}$ and $\vsini = \FitVsiniRMReloaded$~\kms.
When combined with our rotation period measurement from \Tierras, we confirm our earlier conclusion for a polar orbit, with a true obliquity of ${\psi = \FitPsiRMReloaded}$.
Given that the canonical and reloaded RM analyses agree, we choose to adopt the result from the reloaded RM measurement in the remainder of this paper, since it directly utilizes the full shape of the distortion to the stellar spectrum caused by the planet's shadow rather than its effect on the CCF centroids.

\section{Discussion \label{sec:discussion}}

\subsection{The Polar Orbit of TOI-2374\,b \label{ssec:toi2374_orbit}}

Our measurement of the three-dimensional obliquity for TOI-2374\,b identifies it as part of the recently identified population of ``polar Neptunes''.
\citet{Albrecht2021} first identified a potential bimodality in the distribution of true obliquities, 
with measured $\psi$ values divided into aligned ($\psi \approx 0^\circ$) and polar ($\psi \approx 90^\circ$) components.
However, later Bayesian statistical analyses by \citet{Siegel2023} and \citet{Dong2023} found that this feature may not be statistically significant when analyzing the entire sample of measured stellar obliquities.
\citet{Knudstrup2024} and \citet{EspinozaRetamal2024} instead suggested that this preponderance of polar orbits may be specific to sub-Saturn planets with masses $\Mp < 0.3\,\Mjup$ (and possibly also hot Jupiters around hot stars), although the number of obliquity measurements of such planets is too small to firmly detect the bimodality in this sub-sample.
\citet{Handley2025} also argued that planets with mass ratios of $\Mp/\Mstar \approx 10^{-4}$ are most susceptible to being found in polar orbits; TOI-2374\,b has $\Mp/\Mstar = 2.3\times10^{-4}$ and would fit into this pattern.

One possible explanation for why it is primarily sub-Saturn-mass planets that are observed to be on polar orbits is that more massive planets may be able to realign their host stars if the star has a massive convective envelope \citep{Winn2010,Schlaufman2010}.
\citet{Attia2023} proposed a ``tidal realignment parameter'' $\tau \propto 1/t_\psi$, inversely proportional to the timescale for tidal realignment.
Based on a sample of roughly 200 measured obliquities, those authors suggested that $\tau < 10^{-15}$ as an approximate threshold for which such realignment may have modified the stellar obliquity occurred over the $\sim$~Gyr ages of their studied systems.
For the TOI-2374 system, which at roughly 3 Gyr old has an age typical of known exoplanet systems, we computed $\tau \approx 3\times10^{-16}$, suggesting that the observed spin-orbit misalignment has likely not been significantly modified by tidal forces, as expected for such a low-mass planet.

What formation or migration history could lead to sub-Saturns being on polar orbits? Dynamical mechanisms such as von Ziepel-Lidov-Kozai (ZLK) oscillations, planet-planet scattering, or other secular interactions are commonly invoked to explain both the existence of close-in giant planets as well as their orbital misalignments (see e.g. review by \citealt{Albrecht2022} and references therein).
More recently, \citet{Petrovich2020} proposed that dispersal of the protoplanetary disk can lead to resonance sweeping between an inner Neptune and an outer giant planet, tilting the inner planet onto a misaligned orbit.
This mechanism depends on the mass ratio between the inner and outer planets, which could be an alternative explanation for why the preponderance of polar orbits may be limited to sub-Saturns, instead of or in addition to the effects of tidal realignment.
These dynamical pathways to misalign the planet's orbit all require the presence of an outer perturbing object. 
Existing RV observations of TOI-2374 do not have sufficient baseline to permit effective constraints on the presence of additional bodies in the system with orbital separations of a few AU. Continued RV monitoring, combined with time series astrometry from \textit{Gaia} DR4, could reveal such a previously unknown perturber and its orbital properties, shedding light on the system's formation history.

Given that there is already a known third body in the TOI-2374 system: the bound companion star HD 202663, we investigate whether this stellar companion could 
be responsible for the polar orbit of TOI-2374\,b.
With a projected separation of 3000 AU, the orbit of the stellar binary is so distant that it must have begun with significant initial misalignment with the planetary orbit in order to drive ZLK oscillations \citep{Wu2003,Liu2015}.
We computed $\gamma$, the angle between the relative position and velocity vectors between the two stars \citep{Tokovinin1998,Tokovinin2016} given their \Gaia astrometric measurements. We found that $\gamma = 92.1 \pm 0.1^\circ$, indicating that the stellar binary orbit is almost entirely in the sky plane.
Meanwhile, our RM analysis found that the the planet host TOI-2374 has a stellar inclination of $\istar = 22\pm6^\circ$, i.e. we are observing it close to pole-on, and the equatorial plane is also close to the sky plane.
Thus, if the protoplanetary disk began well-aligned with the stellar equator of TOI-2374, the disk plane would also have been relatively well-aligned with the stellar binary orbital plane.
Such a configuration is not compatible with the requirement of primordial misalignments necessary to initiate ZLK oscillations of TOI-2374\,b.

Another way in which a wide-orbiting stellar companion may modify a planet's orbital plane is through torquing of the protoplanetary disk.
Several authors \citep[e.g.,][]{Behmard2022,Dupuy2022,Christian2022} have suggested that this may be responsible for the observed statistical preference for alignments between planetary orbits and the orbits of stellar companions in the same system.
However, this mechanism would tend to align the protoplanetary disk with the binary orbital plane; whereas the current geometry of the TOI-2374 system has the planet's orbit perpendicular to that of the stellar binary, so this is unlikely to have occurred in this case.
Still, we note that many other close-in giant planets that have been found to be on highly misaligned orbits are also in wide binary systems \citep[e.g.,][]{Hagelberg2023,Giacalone2025}. We therefore cannot completely rule out the possibility that the presence of the stellar companion TOI-2374A had some influence on TOI-2374\,b's orbit, for example through more complex dynamical scenarios like four-body interactions \citep{Yang2025}.

\subsection{Neptune Ridge Obliquities \label{ssec:neptune_ridge}}

\begin{figure*}
\plotone{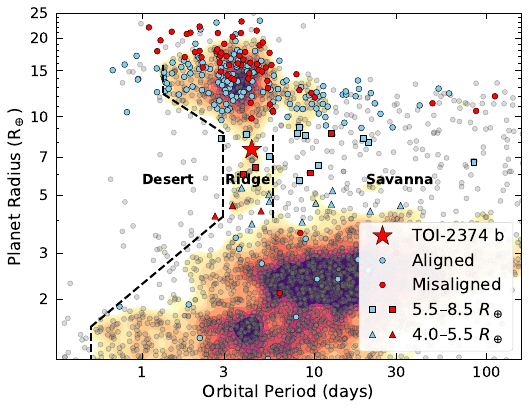}
\caption{Distribution of known transiting planets in the planet radius-orbital period plane, showing the boundaries of the ``hot Neptune desert'' and ``Neptunian ridge'' (the overdensity of intermediate-sized planets between 3 and 6 days between the vertical dashed lines), as defined by \citet{Castro-Gonzalez2024}.
TOI-2374\,b is part of the Neptunian ridge and is shown as the red star.
Planets with no obliquity measurements are shown in gray, planets with low measured obliquities are shown in blue, while planets with misaligned orbits ($\lambda$ more than 1-$\sigma$ greater than 15$^\circ$) are shown in red.
Square symbols denote planets with radii between 5.5 and 8.5\,\Rearth, while triangles denote those with \Rp between 4.0 and 5.5\,\Rearth.
Planet properties were drawn from the NASA Exoplanet Archive \texttt{PSCompPars} table \citep{ExoplanetArchive_PSCompPars} as of 05 Jun 2025, while obliquity measurements were compiled from \citet{Albrecht2022,Knudstrup2024} and the TEPCat catalog \citep{TEPCat_Southworth2011}.
}
\label{fig:nep_desert}
\end{figure*}

With a radius of $0.67\,\Rjup = 7.5\,\Rearth$ and a mass of $0.18\,\Mjup = 60\,\Mearth$, TOI-2374\,b is a sub-Saturn intermediate between the hot Jupiters and the more common sub-Neptunes.
Using data from the \Kepler mission, \citet{Castro-Gonzalez2024} investigated the planet occurrence statistics of planets with sizes in this transition zone, $5.5\,\Rearth < \Rp < 8.5\,\Rearth$.
Those authors found that such planets appear to have a peak in their occurrence rate at $3.2\,\mathrm{days}\lesssim P_\mathrm{orb} \lesssim 5.7\,\mathrm{days}$, which they termed the ``Neptunian ridge'' (Fig. \ref{fig:nep_desert}), reminiscent of the ``three-day pile-up'' for the larger hot Jupiters \citep[e.g.,][]{Udry2003,Gaudi2005}.
\citet{Vissapragada2025} found that the planets in the Neptunian ridge, as well as those on even closer orbits within the ``hot Neptune desert'', are preferentially found around metal-rich stars, a trait shared again with the hot Jupiters but not with longer-period planets of the same size.
These similarities with the hot Jupiters suggest that they may share a common formation pathway, and that the planets in the hot Neptune desert and ridge may even be the product of hot Jupiters that have undergone or are undergoing significant mass loss \citep[e.g.,][]{Ehrenreich2015}.

The orbital properties of these intermediate-sized planets may also provide clues as to whether they formed similarly to the hot Jupiters. \citet{Correia2020} noted that hot Neptunes appear to have non-zero eccentricities, perhaps as the remnant of recent high-eccentricity migration from a more distant orbit.
\citet{Bourrier2023} investigated the obliquity distribution of planets in and around the hot Neptune desert, finding a high fraction of misaligned orbits, consistent with a dynamically violent migration history for these planets.
More recently, \citet{Doyle2025b} also explored the planets in the Neptunian desert, using the boundaries of \citet{Mazeh2016}.
Using six Neptune desert and ridge planets with obliquity measurements from their ``gold'' sample, those authors found no strong evidence for either obliquity or eccentricity excitation, calling into question the importance of the high-eccentricity migration pathway.

With the addition of a few new stellar obliquity measurements since those previous studies, we revisit this population of planets.
We limit our focus specifically to planets in the Neptunian ridge and desert within the boundaries defined by \citet{Castro-Gonzalez2024}: $P_\mathrm{orb} < 5.7$~days, $5.5\,\Rearth < \Rp < 8.5\,\Rearth$. Including TOI-2374\,b, only seven such planets have published obliquity measurements. Of these, three are on polar orbits (WASP-156\,b \citep{Bourrier2023},\footnote{We note that \citet{Bourrier2023} reported that the observations of WASP-156\,b may have been contaminated by telluric features, and \citet{Doyle2025b} recently reported that this planet's orbit may in fact be well-aligned with the stellar spin, based on as-yet-unpublished data. While we adopt the published value, we caution that if WASP-156\,b is not on a polar orbit, this would weaken our statistical conclusions in the strict sample. However, with the relaxed planet radius cut, we still find $p < 0.001$ to find 5/13 polar hot Neptunes.}
TOI-3884\,b \citep{Libby-Roberts2023,Mori2025,Tamburo2025b}, and TOI-2374\,b), while the others have low obliquities consistent with zero.
In contrast, only one out of the nine planets with similar sizes but on longer orbital periods ($P_\mathrm{orb} > 5.7$~days) and an obliquity measurement, has $\lambda$ more than 1$\sigma$ greater than $15^\circ$.
Simple binomial statistics would suggest only a $3\%$ probability that we would have found $\geq3/7$ of hot Neptunes to be on polar orbits if the base rate was 1/9, but we stress that both samples are small and include stars with different spectral types and multiplicity properties.
This result differs from the conclusions of \citet{Doyle2025b} due to the inclusion of two polar planets, TOI-2374\,b and TOI-3884\,b, which were not in their sample.

We also repeated our analysis relaxing the lower radius limit to $\Rp > 4.0\,\Rearth$.
This expands the sample to include six new planets, three of which are on polar orbits: HAT-P-11\,b \citep{Winn2010b,Hirano2011}, GJ 3470\,b \citep{rmfit_Stefansson2022}, GJ 436\,b \citep{Bourrier2022}, while the other three have well-aligned orbits: TOI-942\,b \citep{Wirth2021} K2-33\,b \citep{Hirano2024}, and TOI-5126\,b \citep{Radzom2024}.
The comparison, longer-period ``warm Neptune'' sample also increases by six planets, all of which are consistent with having low obliquities.
Binomial statistics then suggests a probability of just $10^{-4}$ to find 6/13 polar hot Neptunes assuming a base rate of 1/15 (the fraction of misaligned warm Neptunes with $4.0\,\Rearth < \Rp < 8.5\,\Rearth$), providing further evidence that Neptunian ridge and desert planets may be preferentially misaligned compared with their longer-period counterparts.
Future work could make this conclusion more statistically robust with a larger sample as well as accounting for the heterogeneous uncertainties of each obliquity measurement.

Despite the limitations of the current small sample, it is tempting to draw parallels between the Neptunes and Jupiters.
We showed that the shortest period hot Neptunes may have a higher rate of spin-orbit misalignments than those at longer periods.
A similar pattern is also seen for the Jupiter-sized planets, where
the closest hot Jupiters around hot stars (which are unlikely to have caused tidal spin-orbit realignment) have high obliquities, but the more distant warm Jupiters typically have low obliquities \citep{Wang2024}.
One plausible explanation is that two pathways operate to produce hot and warm Jupiters, with one that excites their inclinations only producing those hot Jupiters that are part of the ``three-day pile-up''.
If the same pattern holds for the Neptune-sized planets, perhaps this high-inclination pathway is also responsible for delivering planets to the Neptunian ridge, and the short-period peak for both Neptunes and Jupiters may have a common origin.

\section{Conclusion}

We have presented a stellar obliquity measurement for the TOI-2374 system, which hosts a planet in the Neptunian ridge.
We firmly detected the Rossiter-McLaughlin effect using the KPF spectrograph, measuring a sky-projected obliquity of $\lambda={\FitLambda}$ using a canonical RM analysis and $\lambda = {\FitLambdaRMReloaded}$ with a reloaded RM analysis, the latter of which we treat as our adopted value.
We also measured the planet host star's rotation period to be $\Prot={\FitProt}$~days, which when combined with the projected stellar rotation velocity of $\vsini=\FitVsiniRMReloaded\,\kms$ measured from the RM effect, indicates that we are observing the star nearly pole-on, along its rotation axis. The planet has an orbit that is nearly polar relative to the stellar spin, with a three-dimensional obliquity of $\psi={\FitPsiRMReloaded}$.
This adds to the small but growing number of Neptunian ridge and desert planets with obliquity measurements; these planets appear to have a high rate of significant spin-orbit misalignments that could indicate a high-eccentricity, high-inclination formation pathway common with the hot Jupiters.
The ongoing discovery of transiting hot Neptunes from TESS \citep[e.g.,][]{Naponeillo2025}, along with extreme precision RV instruments like KPF, will continue to grow this sample and potentially reveal the origin of this unusual planet population.

\begin{acknowledgments}
The authors gratefully acknowledge the anonymous reviewer for constructive feedback that greatly improved the manuscript.

S.W.Y. thanks Ale Hacker for providing HARPS activity indicator measurements from the discovery paper.

S.W.Y. and J.G.-M. gratefully acknowledge support from the Heising-Simons Foundation. J.G-M. acknowledges support from the Pappalardo family through the MIT Pappalardo Fellowship in Physics.
Funding for K.B. was provided by the European Union (ERC AdG SUBSTELLAR, GA 101054354).

The \textit{Tierras} Observatory is supported by the National Science Foundation under Award No. AST-2308043. \textit{Tierras} is located within the Fred Lawrence Whipple Observatory; we thank all the staff there who help maintain this facility.

This work was supported by a NASA Keck PI Data Award, administered by the NASA Exoplanet Science Institute. Data presented herein were obtained at the W. M. Keck Observatory from telescope time allocated to the National Aeronautics and Space Administration through the agency's scientific partnership with the California Institute of Technology and the University of California. The Observatory was made possible by the generous financial support of the W. M. Keck Foundation. The authors wish to recognize and acknowledge the very significant cultural role and reverence that the summit of Maunakea has always had within the indigenous Hawaiian community. We are most fortunate to have the opportunity to conduct observations from this mountain.

This paper is based on observations made with the MuSCAT3 instrument, developed by the Astrobiology Center (ABC) in Japan, the University of Tokyo, and Las Cumbres Observatory (LCOGT). MuSCAT3 was developed with financial support by JSPS KAKENHI (JP18H05439) and JST PRESTO (JPMJPR1775), and is located at the Faulkes Telescope North on Maui, HI (USA), operated by LCOGT. This work is partly supported by JSPS KAKENHI Grant Numbers JP24H00017, JP24K00689, and JSPS Bilateral Program Number JPJSBP120249910.

\end{acknowledgments}




%
\facilities{Keck:I (KPF),
\textit{Tierras} Observatory,
FTN (MuSCAT3),
TESS
}


\software{\texttt{astropy} \citep{Astropy13,Astropy18,Astropy2022},
          \texttt{AstroImageJ} \citet{AstroImageJ_Collins17}
          \texttt{numpy} \citep{Numpy},
          \texttt{scipy} \citep{Scipy},
          \texttt{matplotlib} \citep{Matplotlib},
          \texttt{rmfit} \citep{rmfit_Stefansson2022},
          \texttt{nep-des} \citep{Castro-Gonzalez2024}\footnote{\url{https://github.com/castro-gzlz/nep-des}}}






\bibliography{main,instruments,catalogs,software}{}
\bibliographystyle{aasjournalv7}



\end{document}